\documentclass[a4paper]{spie}  

\usepackage[english]{babel}
\usepackage{times}

\usepackage[
letterpaper,
left=0.88in,
right=0.88in,
top=1in,
bottom=1.25in
]{geometry}

\usepackage{amsmath}
\usepackage{graphicx}
\usepackage[colorlinks=true, allcolors=blue]{hyperref}
\usepackage{authblk}
\usepackage{enumitem}
\usepackage{siunitx}
\usepackage{tabularx}
\usepackage{booktabs}
\usepackage{subcaption}
\usepackage{amssymb}
\usepackage{booktabs} 
\usepackage{tabularx}

\title{Overview and design optimization of a custom hybrid X-ray telescope for the International Axion Observatory (IAXO)}

\author[1,*]{Yue Yu}
\author[1]{Jooyun Woo}
\author[1]{Ipek Altunyurt}
\author[2]{Stefano Basso}
\author[3]{Vadim Burwitz}
\author[4]{Francisco R. Candón}
\author[5]{Juan Francisco Castel Pablo}
\author[2]{Marta M. Civitani}
\author[1]{Todd Decker}
\author[6]{Klaus Desch}
\author[7]{Desiree Della Monica Ferreira}
\author[5]{Javier Galan}
\author[5]{Maurizio Giannotti}
\author[8]{Louis Helary}
\author[7,${\dagger}$]{Peter L. Henriksen}
\author[4,5]{Alba María Huerva}
\author[5]{Igor G. Irastorza}
\author[1]{Eftychia Kotsiou}
\author[8]{Axel Lindner}
\author[8]{Frank Marutzky}
\author[7]{Sonny Massahi}
\author[9]{Sam Moseley}
\author[9]{Takashi Okajima}
\author[6]{Johanna von Oy}
\author[7]{Diego Paredes-Sanz}
\author[2]{Giancarlo Parodi}
\author[1]{Kerstin Perez}
\author[10]{Michael J. Pivovaroff}
\author[8]{David Reuther}
\author[1]{Anacorina Romero}
\author[4,10]{Jaime Ruz}
\author[1,11]{Vyshnavi Sabbi}
\author[6]{Tobias Schiffer}
\author[6]{Uwe Schneekloth}
\author[2]{Daniele Spiga}
\author[1]{Marcela Stern}
\author[4,10]{Julia K. Vogel}

\affil[1]{Department of Physics, Columbia University, New York, NY 10027, USA}
\affil[2]{INAF, Italian National Institute for Astrophysics, Osservatorio Astronomico di Brera,
Milano/Merate, Italy}
\affil[3]{Max-Planck-Institut für extraterrestrische Physik, Garching D-85748, Germany}
\affil[4]{Fakultät für Physik, Technische Universität Dortmund, Dortmund, D-44221, Germany}
\affil[5]{Centro de Astropartículas y Física de Altas Energías (CAPA), Universidad de Zaragoza, 50009, Zaragoza, Spain}
\affil[6]{Physikalisches Institut der Universität Bonn, Nussallee 12, 53115, Bonn, Germany}
\affil[7]{Astrophysics and Atmospheric Physics, National Space Institute, Technical University of Denmark, 2800 Kgs. Lyngby Denmark}
\affil[8]{Deutsches Elektronen-Synchrotron DESY, Notkestr. 85, 22607, Hamburg, Germany}
\affil[9]{NASA Goddard Space Flight Center, Greenbelt, MD 20771, USA}
\affil[10]{Physics and Life Science Directorate, Lawrence Livermore National Laboratory, Livermore, CA 94550, USA}
\affil[11]{Department of Physics and Astronomy, The University of North Carolina at Chapel Hill, Chapel Hill, NC 27599, USA}

\authorinfo{\textsuperscript{*}Corresponding author: Yue Yu, yy3571@columbia.edu \\ \textsuperscript{$\dagger$}Present affiliation: Terma A/S, Hovmarken 4, 8520 Lystrup, Denmark}

\begin{document}
\date{}
\maketitle

\begin{abstract}

We present the design optimization for maximizing the effective area of a custom X-ray optic for the International Axion Observatory (IAXO) and BabyIAXO, including its novel hybrid configuration that enables full coverage of the 700-mm-diameter magnetic bore with minimal stress imposed on the mirrors; shell layout optimized for axion spectra and spatial distribution; and the coating recipes that enhance reflectivity in the energy range of interest. We evaluate how these design choices improve the observation signal-to-noise ratio (SNR) of BabyIAXO and IAXO by calculating the broad-band effective area and simulating the point spread function (PSF) and focal spot at the detector plane. 
The cost-effective and scalable optic offers an energy response from 0.03--15\,keV, achieving an effective area that exceeds 2400~cm$^2$ near 1\,keV -- the peak of the ABC axion spectrum -- and remains above $1700\,\text{cm}^2$ around 3\,keV -- the peak of the Primakoff axion spectrum. It yields a half-power diameter (HPD) of $\sim 46^{\prime\prime}$ for an on-axis point source at infinity, and a focal-spot HPD of $\sim 120^{\prime\prime}$ for the radial distribution expected for axion signals within the approximately $3^{\prime}$-radius solar core. 
A relatively generous fabrication-error budget is also summarized. 
The custom optic, accounting for fabrication errors, is anticipated to deliver a more than $55$-fold enhancement in the SNR.

\end{abstract}

\keywords{X-ray, Telescope, IAXO/BabyIAXO, Optical design, Axion}

\section{\label{sec:intro} Solar axions and the next generation of helioscopes}\label{sec:axion-spectra}

Axions and axion-like particles (ALPs) are among the most promising dark matter candidates \cite{Adams:2022pbo, DiLuzio:2020wdo}. Our Sun is a natural axion factory, producing axions through multiple channels \cite{Carenza:2024ehj}. The existence of these diverse production channels makes detection sensitivity across a broad energy range critical, spanning from soft sub-keV components up to the $14.4\,\mathrm{keV}$ nuclear line \cite{Zioutas:2004hi, CAST_aee, Di_Luzio_2022, plasmons}. 
Among these channels, Primakoff production, controlled by the axion–photon coupling ($g_{a \gamma}$), arises from the conversion of thermal photons into axions in the screened Coulomb fields of charged particles in the solar plasma and plays a central role in both solar axion production and helioscope detection. 

Helioscopes, as a powerful and model-independent approach, offer high sensitivity over a broad mass range when probing axions and ALPs from the Sun. The International Axion Observatory (IAXO) \cite{Irastorza:2011gs,BabyIAXOCDR,IAXO:2019mpb} is a next-generation helioscope that reconverts solar axions into X-ray photons in a strong laboratory magnetic field. IAXO will improve sensitivity to the axion–photon coupling $g_{a\gamma}$ by a factor of 20 over the current leading helioscope, the CERN Axion Solar Telescope (CAST~  \cite{Arik:2015cjv,anastassopoulos2017new}), across a wide mass range extending up to $\sim0.25$~eV/$c^{2}$. BabyIAXO, a preliminary pathfinder of IAXO, is designed to validate all key technologies at scale while also achieving an improvement in sensitivity by a factor of $\sim5$ compared to CAST, already opening novel discovery space for axions and ALPs. 

BabyIAXO will use a superconducting magnet with two 700 mm–diameter bores to maximize the collecting area for rare signal searches. Previous study of the IAXO figure of merit \cite{Irastorza:2011gs}, as well as experience with the CAST prototype optic \cite{pivovaroff2026pathfinder}, has proven that the focusing X-ray optics are essential to reduce the spot size and improve the signal-to-noise  (SNR). To fully leverage the photon flux converted by the large magnet, the telescopes will fully cover the bore apertures. 

A custom optic, as one of the BabyIAXO optics, has been designed and is under fabrication as a baseline optic and a pathfinder for IAXO; therefore, this optic employs scalable technologies identical to those envisioned for IAXO. This custom optic is dedicated to detecting weak solar axion signals, implying significant differences in design criteria compared with previous X-ray telescopes.

\begin{figure}[!h]
\centering
\includegraphics[width=120mm]{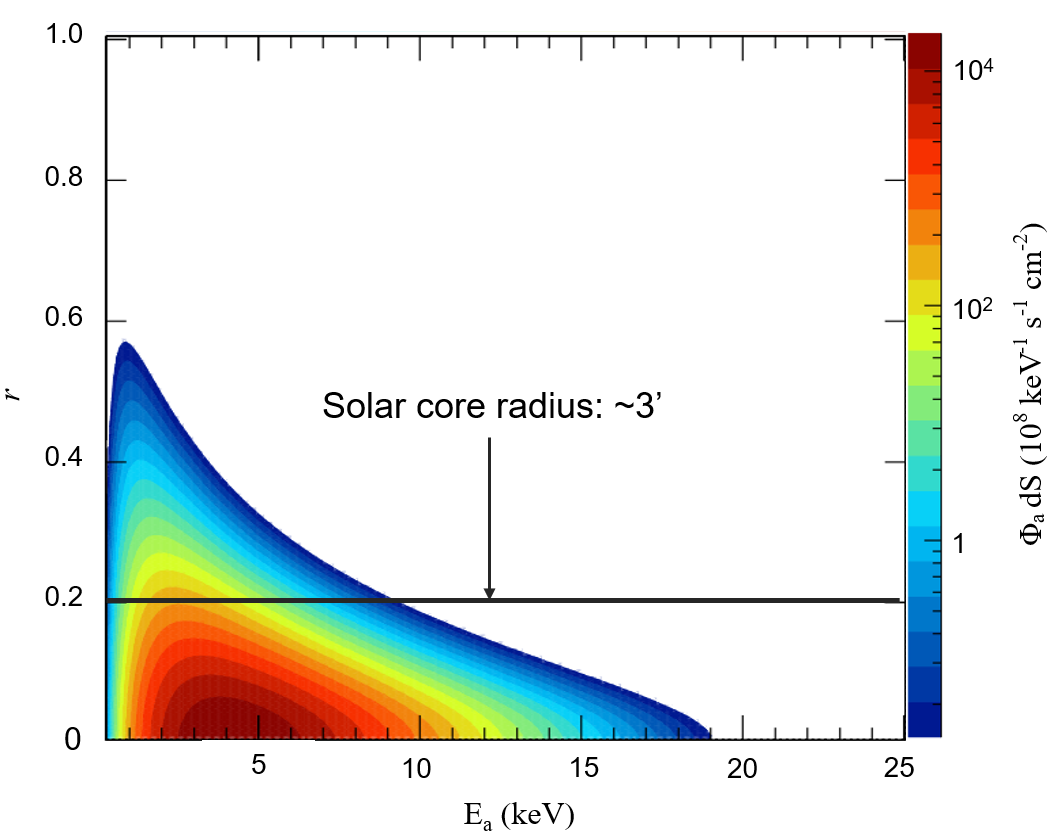}
\caption[Solar axion surface luminosity and differential spectrum]{
      Surface luminosity of solar axions produced via the Primakoff effect, shown as a function of energy and the radial position \(r\) on the solar disk~\cite{CAST2}. The flux is shown for $g_{a \gamma}=10^{-10} \mathrm{GeV}^{-1}$ and scales as $g_{a \gamma}^2$.}
\label{beauty}
\end{figure}

First, the axion signal is expected to originate predominantly from the solar core with known spectra and spatial distribution.
As shown in Figure \ref{beauty}, the surface luminosity distribution of the Primakoff axion flux is highly concentrated within $r \lesssim 0.2$, where $r = R/R_{\odot}$ represents the dimensionless radial coordinate, corresponding to an angular radius of approximately $3^\prime$.
The objective is to collect as many converted X-ray photons as possible and focus them onto the smallest possible detector area, thereby improving the separation of the signal from the detector background. Consequently, the optic design, tailored to the axion spectra and spatial distribution, prioritizes a large effective area (EA) from sub-keV up to beyond 14.4\,keV to maximize the detection sensitivity of weak axion signals across multiple production channels.

To increase EA, the following challenges must be addressed:
\begin{enumerate}[label=(\arabic*)]
    \item Enlarging the diameter to fully cover the large magnetic bore.
    \item Optimizing the optical geometry for axion spectra and spatial distribution.
    \item Maximizing reflectivity across a wide energy range and a broad range of incident angles.
\end{enumerate}

Second, because solar axions reach the Earth through the atmosphere, the entire system--including the optic--can operate on the ground. As a result, IAXO is subject only to gravitational loads and does not need to withstand the launch process, simplifying material selection and environmental testing. Ultra-low-background measurements also require radiopurity of materials, since the optic will be positioned within the same vacuum vessel as the detectors. Finally, as a pathfinder for IAXO, which will require eight custom X-ray optics, low-cost and scalable design are essential.

To satisfy the above criteria, the custom BabyIAXO optic adopts segmented glass techniques~\cite{koglin2004production,Koglin2003,Koglin2005,Civitani2013oe,civitani2016cold}. This kind of techniques allows flexible coating designs to enhance broadband X-ray reflectivity, and the shells can be nested tightly to maximize the collecting area while maintaining modest spatial resolution in a cost-efficient way. In its final phase of operation, CAST hosted a small (EA of a few cm$^{2}$) custom X-ray optic sector, demonstrating the feasibility of incorporating focusing X-ray optics utilizing segmented glass technologies to improve axion sensitivity \cite{Aznar:2015iia,adair2022search, pivovaroff2026pathfinder,jakobsen2015x,10.1117/12.2024476}. Previous works summarized the conceptual designs of IAXO~\cite{Irastorza:2011gs} and BabyIAXO~\cite{BabyIAXOCDR} and developed the reflective coating recipes optimized for axion-induced spectra~\cite{Henriksen:21}.

\section{\label{sec:design} Design optimization for a hybrid optic}
\subsection{Hybrid structure for large-diameter coverage}
\begin{figure*}[ht]
\begin{centering}
\includegraphics[width=1\linewidth, angle=0]{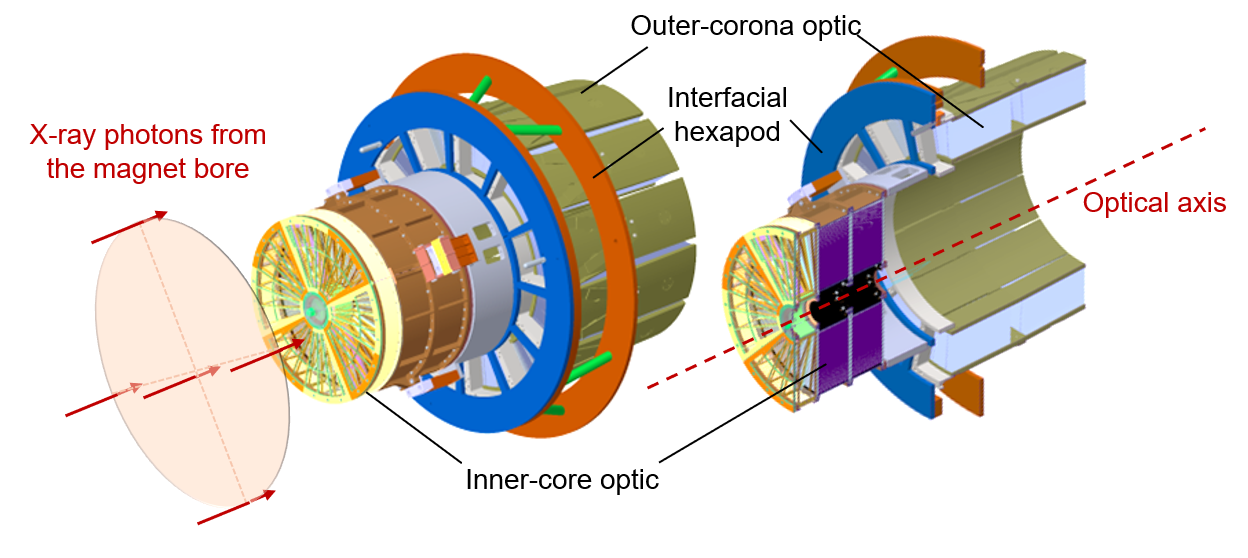}
\vspace{-0.in}
\caption{\label{fig:hybrid} Hybrid optic mounting concept. The left panel provides an overview of the structure and its orientation toward the magnet bore, while the right panel shows a side cross-section illustrating their relative position along the optic axis.}
\end{centering}
\end{figure*}

The custom telescope will comprise nested shells of thin glass mirror segments with radii ranging from 54\,mm to 350\,mm to cover the entire magnet bore.
To our knowledge, this will be the largest-diameter segmented thin-glass X-ray telescope yet deployed. Covering such a large radial span is a technical challenge in itself.

Both the thermally slumped glass optics (TGO) technique and the cold glass slumping optics (CGSO) technique are well-developed segmented-glass approaches. The TGO technique has already demonstrated excellent performance over a wide range of shell radii in the HEFT~\cite{Koglin2003b,koglin2004production} and NuSTAR~\cite{koglin2011first,hailey2010nuclear} projects, where thermally slumped and coated mirror segments were assembled from the inner radius outward by stacking mirror layers onto spacers. Since the current optic requires shell radii up to 350\,mm, substantial shear stress may potentially be imposed on the inner shells when the shell radii span such a wide range. The CGSO technique, developed and patented (TO2015A000219) by INAF/OABrera~\cite{Civitani2013oe,civitani2016cold}, on the other hand, is specifically tailored for producing lightweight, large-radius grazing-incidence X-ray mirrors by directly integrating coated flat glass foils into X-ray optic units (XOUs).

To take advantage of both techniques and to mitigate potential adhesive failure or mirror deformation, a hybrid structure is adopted~\cite{BabyIAXOCDR}. The hybrid optic consists of an outer-corona extension and an inner-core optic, as shown in Fig.~\ref{fig:hybrid}. They share a common axis and vacuum vessel but have different focal lengths. The outer-corona extension uses the CGSO technique for large-radius shells, while the inner-core optic employs the TGO technique by leveraging the facilities and equipment used for the HEFT~\cite{Koglin2003b,koglin2004production}, NuSTAR~\cite{koglin2011first,hailey2010nuclear}, and CAST pathfinder~\cite{pivovaroff2026pathfinder} optics. A focal length of 5.0--5.2\,m was initially determined as an optimal compromise between collection area and focal-spot size~\cite{Henriksen:21,BabyIAXOCDR}. Mechanical constraints led to a slight modification of the actual focal lengths, with the inner-core optic set to 5.6\,m and the outer-corona extension set to 5.0\,m.

The outer-corona shells follow a quasi-Wolter-I geometry, with a single tandem module used for a XOU. In each XOU, only the central shell follows a true Wolter-I geometry, while the other shells share the same nominal shape as the central shell but have their tilt angles tuned to achieve improved focusing.
Meanwhile, the inner-core optic adopts a conical approximation to the Wolter-I geometry. Both configurations exhibit intrinsic design aberrations of a few tens of arcseconds, but these errors have a negligible impact on the focal-spot size, which is dominated by the angular extent of the solar axion-emitting region. Practical errors arising from mirror-segment fabrication are expected for both configurations; nevertheless, the resulting degradation in SNR enhancement is negligible, making both approaches cost-effective solutions for axion detection.

The major challenges in realizing a hybrid coaxial structure include: (1) aligning the inner and outer optics either independently or as a rigid, coupled assembly inside the vacuum tube; (2) achieving fine co-alignment to maintain their relative distance and ensure that their focal spots coincide on the detector; and (3) allowing the telescope and mounts to be operated outside the vacuum to minimize the volume and cost of the vacuum vessel and improve serviceability. Operation in a magnetic field of at least 2\,T also increases the complexity of the drive electronics and the overall mechanical design.

A hexapod design, as shown in Fig. \ref{fig:hybrid}, in which a hexapod is mounted externally on the vacuum tube and connected to the telescope support structure via bellows feedthroughs, has been developed to provide the required functionality with controlled cost and without mechanical interference. This configuration preserves the full adjustment capability of an in-vacuum hexapod while avoiding in-vacuum serviceability and outgassing risks.

\subsection{Coating recipe optimization}

A large effective area (EA) is crucial for a rare signal search experiments such as IAXO, and high-reflectivity coatings are the key to maximizing the EA. 
We optimize the radially grouped reflectivity over the energy range 0.03--15\,keV. The lower bound is chosen not only to cover the majority of the spectrum relevant to plasmon-axion~\cite{plasmons} detection, but also because 0.03 keV is close to the lower-energy limit of the Henke optical-constants database \cite{cxro}. The upper bound of 15 keV allows the optic to remain applicable to axion-nucleon coupling~\cite{CAST1, Di_Luzio_2022}.
High-Z materials such as platinum, iridium, and tungsten are natural candidates due to their larger critical angles in the target energy range, particularly from 0.5 to 10 keV. Over this energy range, multilayer coatings are not expected to provide a substantial advantage over single-layer or bilayer coatings in current geometry, given the much higher coating-time commitment and target cost of multilayer coatings. A single-layer coating of high-Z materials, however, would have a sharp reflectivity drop near the M absorption edge. This can be improved by adding a light-material (e.g., C or B$_4$C) top layer. The bilayer coating, therefore, enhances the reflectivity at lower energies. A study \cite{Henriksen:21} presented preliminary optimization of a Ir/C bilayer recipe. Due to the availability and cost of target materials, and based on experience acquired from NuSTAR~\cite{10.1117/12.894615}, we updated Pt/C as the bilayer material pair for coating recipe optimization. 

\begin{figure*}[htbp]
  \centering
  
\begin{minipage}[c]{0.48\linewidth}
    \centering
    \captionof{table}{Custom Pt/C bilayer coating recipes.}
    \label{tab:coatings}
    \vspace{5pt} 
    \setlength{\tabcolsep}{3pt}
    \begin{tabular}{ccccc}
      \toprule
      Recipe & Radius (mm) & $\alpha$ (deg) & Pt (nm) & C (nm) \\ 
      \midrule
      1 & 300--350 & 0.86--1.00 & 10.0 & 5.6 \\ 
      2 & 250--300 & 0.71--0.86 & 10.0 & 7.9 \\ 
      3 & 200--250 & 0.48--0.71 & 10.0 & 10.1 \\ 
      4 & 150--200 & 0.36--0.48 & 10.0 & 13.5 \\ 
      5 &  54--150 & 0.15--0.36 & 10.0 & 18.0 \\ 
      \bottomrule
    \end{tabular}
  \end{minipage}
  \hfill 
  \begin{minipage}[c]{0.5\linewidth}
    \centering
    \includegraphics[width=1\linewidth]{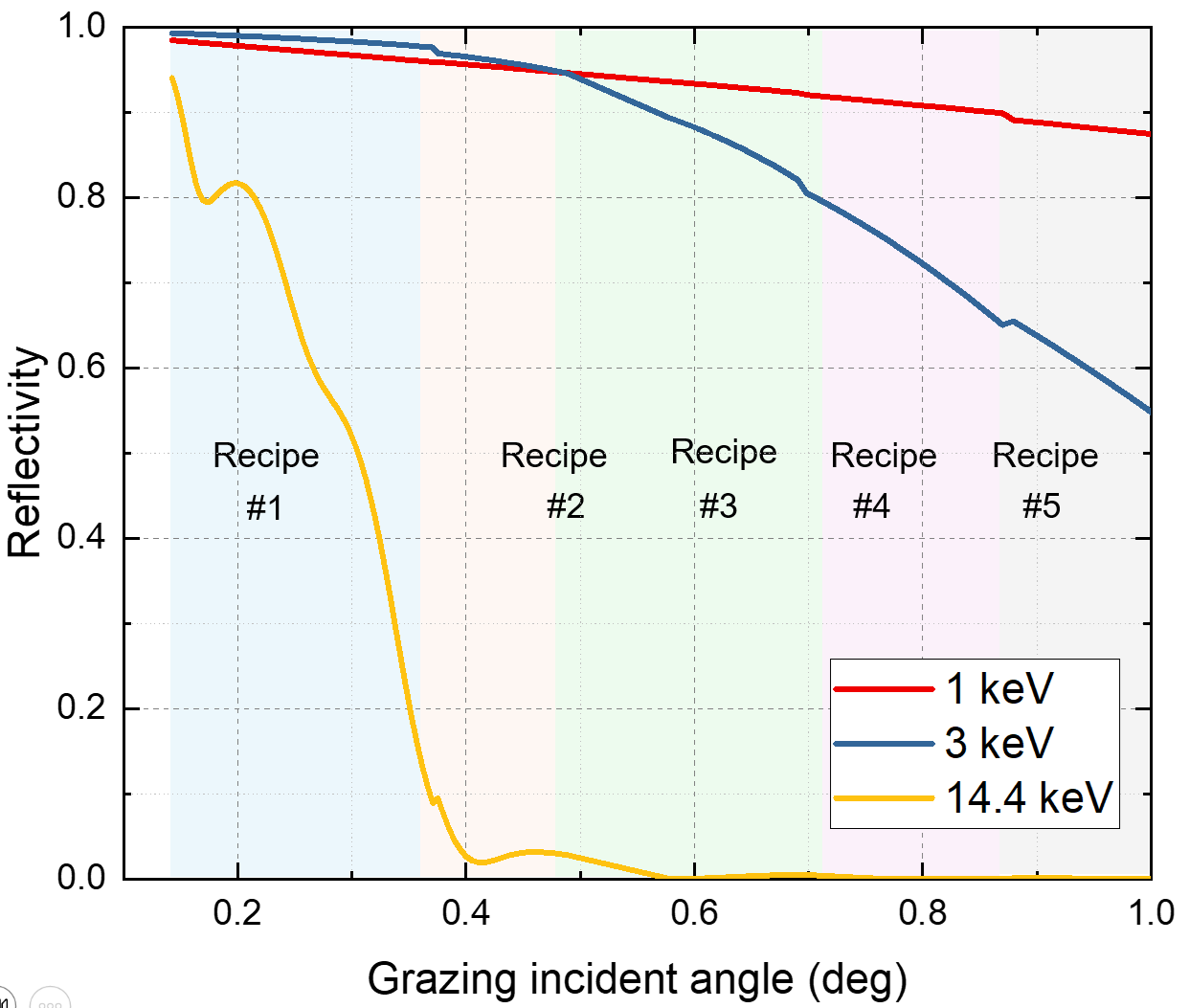}
    \caption{Reflectivity performance of the combined coating recipes at 1\ keV, 3\ keV and 14.4\ keV as a function of grazing incident angle.}
    \label{fig:R}
  \end{minipage}

\end{figure*}

The shells are divided into five subgroups according to their diameter, and the Pt/C coating thicknesses are optimized using a Markov Chain Monte Carlo (MCMC) procedure, as listed in Table \ref{tab:coatings}. A 10 nm Pt thickness is sufficiently large that further increases have a negligible impact on reflectivity within the optimization energy range. Taking into account the deposition-induced roughness, the thickness of the Pt layer is therefore fixed at 10 nm for all groups. The reflectivity is, instead, more sensitive to the thickness of the C-layer. For the outer shells, a thinner C layer enhances the effective area at low energies while suppressing it at higher energies; In contrast, for the inner shells, a thicker C layer improves the effective area at medium to high energies within the optimization band. Consequently, the optimized C layer thickness increases toward the inner shells.
The performance of the combined coating recipes at representative energies is presented in Fig.~\ref{fig:R} as a function of grazing-incidence angle, with colored bands marking the angular range covered by each recipe. The reflectivity of the bilayer coating is sensitive to the incidence angle, particularly at the high-energy end, where it decreases significantly with increasing grazing incidence angle.

\subsection{Vignetting effect analysis and geometry optimization}

Unlike telescopes optimized for large field-of-view (FOV) astrophysical observations, the telescope in this work aims to receive axion-converted X-ray photons that pass through the magnet and are primarily distributed within a $3'$ radius. For each axion spectral component, their ring-integrated radial flux distribution, obtained by integrating the corresponding axion spectral flux over rings on the solar disk, peaks at different off-axis angles. This requires a dedicated optimization of the spacing between the shells to maximize the X-ray throughput.

\begin{figure*}[ht]
\begin{centering}
\includegraphics[width=1\linewidth, angle=0]{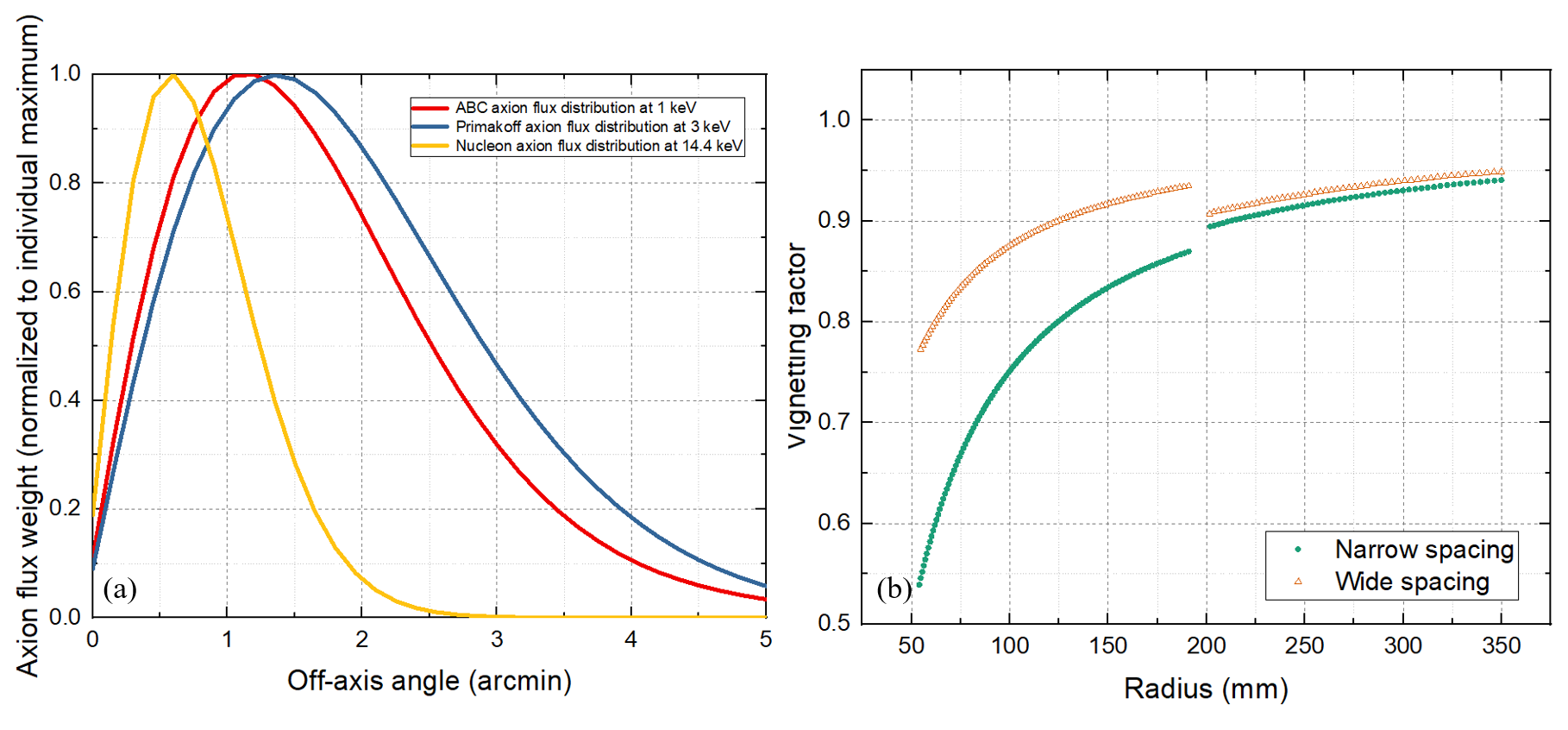}
\vspace{-0.in}
\caption{\label{fig:vignetting} (a) the relative ring-integrated fluxes of ABC axion, Primakoff axion and nucleon-related axion, as a function of off-axis angles; (b) vignetting effect as a function of telescope radius.}
\end{centering}
\end{figure*}

At representative energies of 1\,keV, 3\,keV, and 14.4\,keV, corresponding respectively to the spectral peak or characteristic energies of the ABC axions (produced by the axion-electron coupling), Primakoff axions, and nucleon-related axions~\cite{Zioutas:2004hi, CAST_aee, Di_Luzio_2022}, the weighting factor $w(\theta,E)$ is obtained by normalizing the ring-integrated radial flux distribution to its own maximum, as shown in the Fig.~\ref{fig:vignetting}(a). Since each radial position on the solar disk maps directly onto a corresponding off-axis angle, this normalization assigns the appropriate physical weight to each angle.

We initially adopted a tightly nested structure, in which only the on-axis incident X-rays can fully illuminate each mirror shell without being obstructed by the inner shells, and we refer to this as the ``narrow spacing design".
As observed in the Fig. \ref{fig:vignetting}(b), the vignetting effect becomes much more severe towards the inner shells, obstructing almost half of the overall weighted off-axis X-rays. This motivates increasing the spacing between these shells until the off-axis angles corresponding to the radial flux peak are fully accepted.

For Primakoff axions, the shell spacing is enlarged to accept converted X-rays at an off-axis angle of $3^\prime$ signals without obstruction, which is referred to as the ``wide spacing design".
The vignetting factor increases substantially for the innermost layers and mildly for the outer corona shells. Enlarging the spacing between the shells provides a larger vignetting factor for off-axis angles. However, whether the shell spacing should be enlarged depends on the combined effects of both the reflectivity and the overall geometric collecting area. 

Taking reflectivity variation and off-axis weighting factor into consideration, the effective areas of the narrow-spacing configuration and wide-spacing configuration are compared. The narrow-spacing configuration is found to provide the largest EA for both the inner-core optic and the outer-corona shells, indicating that the gain from accepting additional off-axis signals does not compensate for the loss of near-on-axis signals with high reflectivity.

The structural parameters derived given the above constraints are summarized in Table \ref{tab:geometry}. The design includes 103 layers for the inner-core optic and 54 layers for the outer-corona optic, covering a radius from 54\,mm to 350\,mm.

\begin{table}[h]
    \centering
    \caption{Design parameters of the BabyIAXO custom optic.}
    \label{tab:geometry}
    \begin{tabularx}{\linewidth}{lXX}
        \toprule
        Parameters & Outer-corona optic & Inner-core optic \\
        \midrule
        Innermost radius (mm)          & 193   & 54   \\
        Outermost radius (mm)          & 350   & 191  \\
        Incident angle $\alpha$ (deg) & 0.52--0.96 & 0.15--0.49 \\
        Focal length (m)          & 5.0   & 5.6  \\
        Number of shells           & 54    & 103  \\
        Mirror length (mm)        & 200   & 225 or 2x112  \\
        Substrate type            & Willow glass & Borosilicate glass \\
        Mirror thickness (mm)     & 0.10  & 0.21 \\
        Number of mirrors            & 1296   & 1824 \\
        Interval gap (mm)         & 50     & 4    \\
        \bottomrule
    \end{tabularx}
\end{table}

\section{\label{sec:design} Optical performance evaluation}
\subsection{Effective area evaluation}

With a determined geometry and coating recipes, the effective area $A_{\mathrm{eff}}$ is defined as:

\begin{equation}
A_{\mathrm{eff}}(E)=\sum_{\theta=0}^{\theta_0}\sum_{i=1}^{N}\int_{0}^{\pi/2} dGC A(i,\theta,\varphi)
\times r_1(i,E,\theta,\varphi)\times
r_2(i,E,\theta,\varphi)\times w(\theta,E).
\end{equation}
where $\theta$ denotes the off-axis angle, ranging from 0 to $\theta_{0}$, where $\theta = 0$ corresponds to the on-axis signal from the center of the solar core, and $\theta_{0} = 3'$ corresponds to the edge of the solar core. The index $i$ denotes the shell number. $\varphi$ represents the azimuthal angle around the optical axis. $dGCA$ represents the differential geometric collection area of a mirror slice, which is a function of $\varphi$, $\theta$, and $i$. The \textit{dGCA} calculation accounts for vignetting effects from off-axis angles and obstruction effects from the inner shells, and the analytical expressions are cross-checked with the vignetting model in Ref.\cite{spiga2011optics,spiga2009analytical}. The variables $r_{1}$ and $r_{2}$ represent the reflectivities of the primary and secondary mirrors, respectively, which depend on both the incident angle on the mirror-slice and the photon energy. For on-axis incidence ($\theta=0$), $r_{1} = r_{2}$. The weighting factor $w(\theta, E)$ is applied to the calculation for each value of $\theta$ and E.

\begin{figure*}[ht]
\begin{centering}
\includegraphics[width=1\linewidth, angle=0]{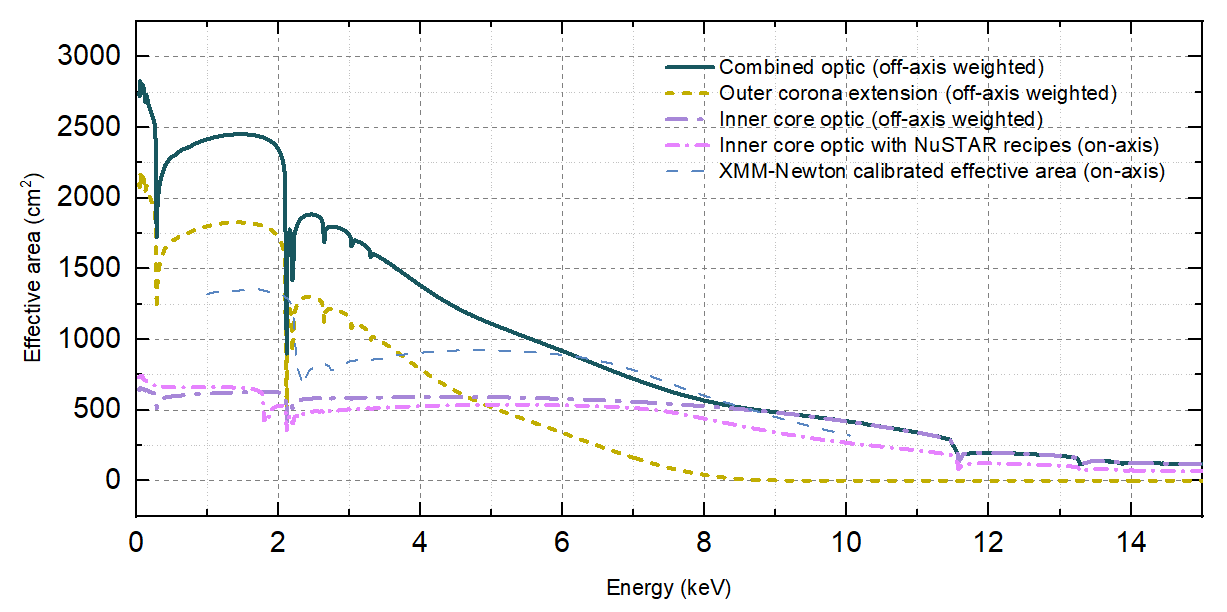}
\vspace{-0.in}
\caption{\label{fig:EA} The simulated off-axis weighted effective area (EA) of the combined optic and the separate contribution from the outer-corona extension and inner-core optic, optimized with five Pt/C bilayer recipes. The simulated on-axis EA of the inner-core optic using NuSTAR heritage coating recipes is overplotted. The calibrated on-axis EA of XMM-Newton flight spare module is also shown as a reference.}
\end{centering}
\end{figure*}

The EA of the combined optic with five optimized bilayer recipes is shown as the bold solid curve in Fig. \ref{fig:EA}. It exceeds 2400 cm$^2$ near 1 keV, which corresponds to the peak of the ABC axion spectrum, and remains beyond 1700 cm$^2$ in the vicinity of 3 keV, close to the peak energy of the Primakoff axion spectrum. The on-axis calibrated EA curve of the spare flight module of XMM-Newton is overplotted as a reference. An EA generally exceeding 2000 cm$^2$ below 1 keV substantially improves the sensitivity to plasmon axion and dark photons \cite{O_Shea_2024}. 
The EA around 14.4\,keV exceeds 100\,cm$^2$ with the current bilayer coating, while a narrow-band coating recipe specifically optimized for 14.4\,keV is being conceived. Such a dedicated coating design, with a bilayer block deposited on top of a multilayer block, could be applied to the innermost few shells to further boost the EA, highlighting the potential sensitivity to axions from nucleon-related channels.

The contributions from the inner-core optic and the outer-corona extension are also shown separately. The outer corona shells dominate the EA at low energies ($ < 5$ keV), whereas the inner-core optic plays a critical role above 5 keV, contributing several hundred cm$^2$ to the EA.

The inner-core optic will utilize both custom-coated mirrors and an inventory of leftover mirrors from the NuSTAR telescope, which used 11 different depth-graded multilayer coating recipes of Pt/C and W/Si \cite{2009NuSTAR}. The on-axis performance of NuSTAR heritage coatings, when applied to the inner-core optic geometry, is also illustrated in the Fig. \ref{fig:EA}. Although the focal length of the BabyIAXO optic is roughly half that of NuSTAR, the associated reflectivity loss is mitigated by the use of depth-graded multilayer coatings. According to Bragg’s law, the high-reflectivity band of the NuSTAR coatings shifts to lower energies under the BabyIAXO geometry, which matches BabyIAXO’s lower-energy requirements while maintaining excellent reflectivity. A slightly lower EA in some parts of the energy range is expected for the mixed configuration employing both NuSTAR heritage and custom coatings; this reduction is acceptable given the gains in time and cost efficiency.

\subsection{PSF and focal spot ray-tracing}

\begin{figure*}[ht]
\begin{centering}
\includegraphics[width=1\linewidth, angle=0]{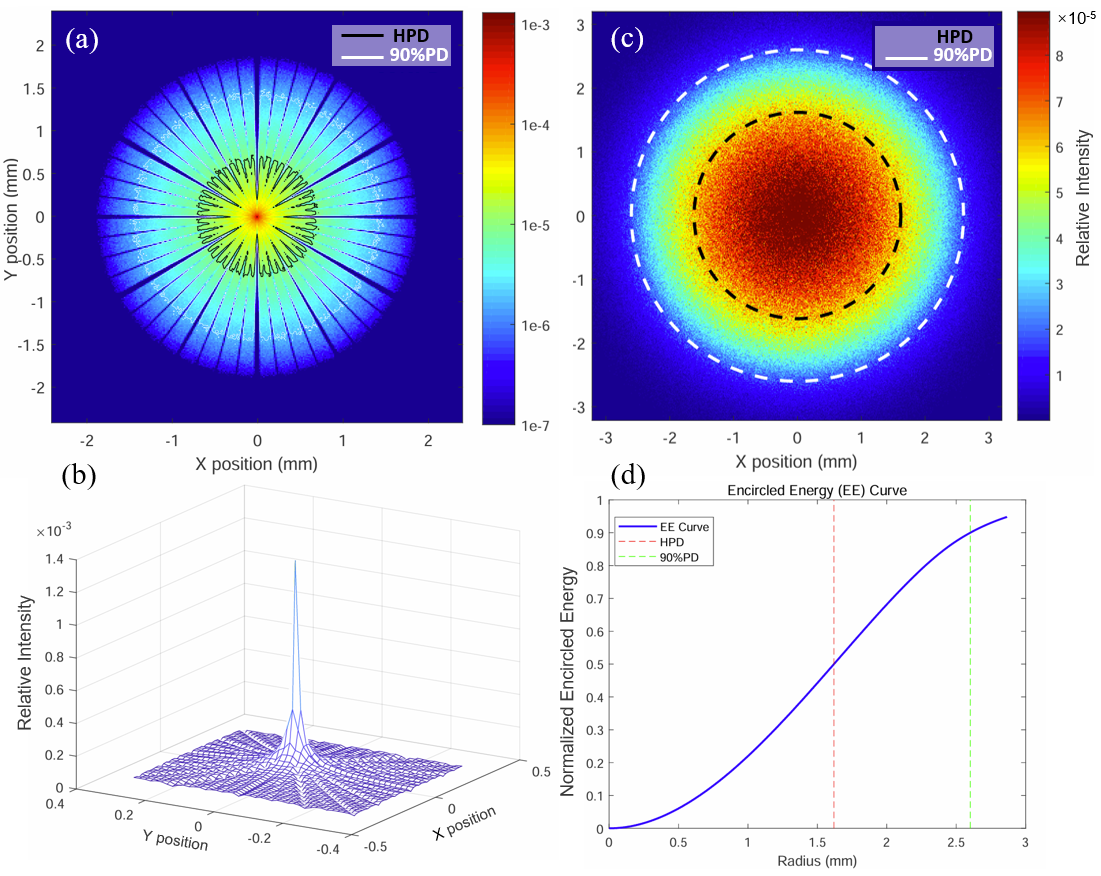}
\vspace{-0.in}
\caption{\label{fig:PSF} (a) Simulated PSF at 3\,keV and (b) its 3D profile. (c) The focal spot images at 3\,keV for the Primakoff axion distribution within $3'$ radius, and (d) its normalized encircled photons as a function of distance from the spot center.}
\end{centering}
\end{figure*}

We performed optical ray tracing using the McXtrace  software\cite{Mcxtrace}, firstly with infinite on-axis point source to evaluate the point spread function (PSF). The detector is placed at the focal plane with a pixel size of 16 × 16 ${\mu}m^2$. The custom bilayer coatings are included to account for the effect of reflectivity on the spot shape. Geometric obstructions, including the mandrel, the spider, and the spacers between shells, are also included in the ray-tracing model, while only the interface mount between inner-core optic and outer-corona shells are not shown here, which will be optimized to minimize the shadowing. Figures \ref{fig:PSF}(a) and (b) show the telescope PSF at 3\,keV in a logarithmic 2D color-map and a 3D contours, respectively. The black and white contours overlaid on the spots represent the 50\% and 90\% encircled energy boundaries, respectively. The simulations yield a PSF half-power diameter (HPD) of approximately 1.25 mm, corresponding to about $46''$. Compared with that of the XMM-Newton flight spare module, which has a calibrated on-axis HPD of only $11.7^{\prime\prime}$ at 58.4\,nm \cite{stockman1999xmm}, the design PSF of the custom optic is admittedly modest. Nevertheless, it remains adequate for our detection goal.

In practice, the angular extent of the solar axion source blurs and widens the focal spot significantly. An equivalent disk source is implemented, with emission points uniformly distributed over the disk, and an off-axis angle initially generated randomly within the range of $[-3', 3']$ for each emitted X-ray. A corresponding intensity weight, $w'(\theta,E)$, is assigned to each X-ray  according to the radial differential Primakoff axion flux, given that the disk source properly accounts for the spatial distribution. Figures \ref{fig:PSF}(c) shows the simulated spots on the focal plane at 3\,keV. Fig. \ref{fig:PSF}(d) displays the normalized encircled energy as a function of the radial distance from the spot center, revealing an angular focal-spot HPD of approximately $120''$ at 3\,keV. 
The HPD values of focal spot vary with photon energy. Toward the higher-energy end, the HPD becomes smaller because both the source flux and the EA are relatively lower. Nevertheless, we use the HPD at 3\,keV to obtain a conservative estimate of the SNR enhancement.

\subsection{SNR enhancement}

The sensitivity of the experiment to the axion coupling constant scales as $N_\gamma/\sqrt{N_b}$, where $N_\gamma$ is proportional to the optical throughput, and $N_b$ is determined by the detector background within the effective spot size \cite{Irastorza:2011gs, pivovaroff2026pathfinder} (e.g., delimited by the HPD in this paper). Therefore, the corresponding SNR improvement $\Delta$SN can be written as:
\begin{equation}
    \mathrm{\Delta{SN}} = \frac{\epsilon_t}{\sqrt{s/A}},
    \label{eq:placeholder_label}
\end{equation}
where $s$ is the effective spot area, and A is the cross-section area of the corresponding magnet bore. $\epsilon_t$ is the total telescope efficiency:
\begin{equation}
    \mathrm{\epsilon_t} = \frac{ \int_{E_1}^{E_2} A_{\mathrm{eff}}(E) \, dE}{GCA_{\mathrm{total}}(E_2-E_1)},
    \label{eq:placeholder_label}
\end{equation}
where $GCA_{\mathrm{total}}$ is the integrated geometric collection area over all shells and is independent of photon energy, whereas the integral term is energy-range dependent. In the present design study, we adopt $E_{1}=0.03~\mathrm{keV}$ and $E_{2}=15~\mathrm{keV}$ for a general evaluation for different axion production channels. By incorporating the HPD of the focal spot and the diameter of the magnet bore, an $\epsilon_t$ of approximately 0.3 is inferred; and assuming that the background performance remains identical with and without optics, we obtain $\Delta \mathrm{SN} > 65$ compared with the case without a focusing telescope. If evaluation is made more specifically for Primakoff axion over a narrower energy range, taking $E_{1}=0.03~\mathrm{keV}$ and $E_{2}=10~\mathrm{keV}$, for instance, $\epsilon_t$ increases to above 0.4 and $\Delta \mathrm{SN}$ reaches $\sim90$.

\section{\label{sec:fabrication} Fabrication error budget}

Figure errors are introduced at each step of the fabrication process. 
Combined with NuSTAR~\cite{koglin2011first, hailey2010nuclear, craig2011fabrication} and HEFT~\cite{Koglin2003b, koglin2004production, Koglin2004} empirical values, and based on the approach established for previous CGSO prototypes~\cite{Civitani2013oe, Basso2015spie}, an error budget for the outer-corona shells and the inner-core optic are listed in Table \ref{tab:errorbudget}. 
The HPD error contributors are categorized into: the intrinsic design aberrations discussed above, target mirror-fabrication figure errors~\cite{Jooyun2026SPIE}, mounting-induced errors, and other second-order error contributors.
Since these error contributions are approximately orthogonal, the total PSF HPDs, accounting for figure errors, are less than $90''$ for both approaches.
Fabrication details are covered in~\cite{Jooyun2026SPIE, Yue2026PRD, Marta2026SPIE}. Even taking the fabrication errors into consideration, a SNR enhancement of $\Delta \mathrm{SN}$$>55$ in the 0.03--15\,keV energy range is expected.

\begin{table}[h]
    \centering
    \caption{Error budget allocation for the inner-core and outer-corona optics.}
    \label{tab:errorbudget}
    \begin{tabular}{lcc}
        \toprule
        \textbf{Error contribution} & \textbf{Inner-core HPD errors ($''$ )} & \textbf{Outer-corona HPD errors ($''$ )} \\
        \midrule
        Design approximation                 & $\sim 46$ & $<20$ \\
        Mirror fabrication                   & $<76$     & $<45$ \\
        Integration errors                   & $\sim 15$ & $<70$ \\
        Scattering/gravity sag/CTE mismatch  & $<5$      & $<2$  \\
        \midrule
        Total PSF HPD ($''$)                    & $<90$     & $<85$ \\
        \bottomrule
    \end{tabular}
\end{table}

\section{\label{sec:conclusion} Summary and outlook}
We have optimized the optical design of the BabyIAXO custom X-ray telescope for the optimal EA to better serve axion observations. The optimized EA exceeds 2400~$\mathrm{cm}^2$ near the peak of the ABC axion spectrum (around 1\,keV) and remains more than 1700~$\mathrm{cm}^2$ near the peak of the Primakoff axion spectrum (around 3\,keV). An EA generally above 2000~$\mathrm{cm}^2$ below 1\,keV further opens sensitivity to plasmon axions and dark photons~\cite{O_Shea_2024}, while $\sim100$\,$\mathrm{cm}^2$ around 14.4\,keV provides potential sensitivity to nucleon-related axion and dark-photon channels~\cite{O_Shea_2024}. 
Without fabrication errors, ray-tracing results show an ideal HPD of $\sim 46^{\prime\prime}$ for an on-axis infinite point source, and $\sim 120^{\prime\prime}$ for the expected axion-signal radial distribution within the $\sim 3^{\prime}$ solar core. We summarized the fabrication error budget. Taking the fabrication errors into consideration, a more than $55$-fold enhancement in SNR is expected with the custom optic.

Assembly of a 10-layer prototype of inner-core optic is planned for mid-2026, and calibration at PANTER is scheduled for later that year. If the measured performance meets our expectations, the 10-layer prototype will serve as the inner 10-layer module of the full telescope, which is expected to finish construction around end of 2027. A proof-of-concept prototype (POC-2023) was realized to validate the CGSO fabrication and integration chain under conditions representative of the BabyIAXO corona design. The POC-2023 was characterized in X-rays at the PANTER facility (MPE, Garching) in 2023. HPD values measured across different azimuthal sectors of the module ranged from $\sim$\,43\,arcsec to $\sim$\,83\,arcsec. 
These results are consistent with the CGSO error budget predictions and confirm that the process delivers optical performance well within the fabrication target. This  provides a solid experimental basis for scaling the technology to the full-aperture XOUs required for the outer corona. Integration and calibration of the first 5-layer XOU prototype of BabyIAXO are scheduled for 2026.

\acknowledgments
We thank Prof. Chuck Hailey for his helpful advice.
This work was supported in part by the National Science Foundation (NSF) PHY-2309980 and NSF Research Experiences for Undergraduates (REU) Program under Grant NSF PHY-2349438. 

JKV, FRC, and JR acknowledge support from the Deutsche Forschungsgemeinschaft (DFG, German Research Foundation) under Germany’s Excellence Strategy – Cluster of Excellence “Color meets Flavor”, EXC 3107 – Project-ID 533766364. JKV and FRC also acknowledge funding from the German federal and state program "Professorinnenprogramm 2030" Project-ID 01FP24167Q.

We acknowledge support from the US Department of Energy (DOE) under Contract No. DE-AC52-07NA27344, performed at the Lawrence Livermore National Laboratory. 
\bibliographystyle{spiebib}
\bibliography{bib}    

\end{document}